\documentclass[12pt]{article}
\usepackage{epsfig,amssymb}

\newcommand{\bq}{\begin{equation}}
\newcommand{\eq}{\end{equation}}
\newcommand{\bqa}{\begin{eqnarray}}
\newcommand{\eqa}{\end{eqnarray}}
\newcommand{\ben}{\begin{enumerate}}
\newcommand{\een}{\end{enumerate}}
\newcommand{\bc}{\begin{center}}
\newcommand{\ec}{\end{center}}
\newcommand{\bqb}{\begin{eqnarray*}}
\newcommand{\eqb}{\end{eqnarray*}}

\def\gsim{\gtrsim}
\def\lsim{\lesssim}

\def\pr#1#2#3{ Phys. Rev. ${\bf{#1}}$ (#2) #3}

\def\pl#1#2#3{ Phys. Lett. ${\bf{#1}}$ (#2) #3}
\def\prep#1#2#3{ Phys. Rep. ${\bf{#1}}$ (#2) #3}

\def\np#1#2#3{ Nucl. Phys. ${\bf{#1}}$ (#2) #3}
\def\zp#1#2#3{ Z. f. Phys. ${\bf{#1}}$ (#2) #3}

\def\ie{{\it i.e.\/}}
\def\eg{{\it e.g.\/}}

\def\etal{{\it et.al.\/}}

\global\nulldelimiterspace = 0pt

\def\L{ {\cal L }}

\def\t{\hat t}
\def\s{\hat s}
\def\u{\hat u}

\begin{document}
\pagenumbering{arabic}
\thispagestyle{empty}
\def\thefootnote{\fnsymbol{footnote}}
\setcounter{footnote}{1}

\begin{flushright}
PM/98-41 \\    THES-TP 98/09 \\
%hep-ph/.....\\
December 1998
 \end{flushright}
\vspace{2cm}
%---------------------titre ---------------------------------------
\begin{center}
{\Large\bf Light by Light Scattering at High Energy:}\\
{\Large\bf a Tool to Reveal New Particles  }\footnote{Partially
supported by the NATO grant CRG 971470 and by the Greek
Government grant PENED/95 K.A. 1795.}
 \vspace{1.5cm}  \\
%-----------------------------------------------------------------
{\large G.J. Gounaris$^a$, P.I. Porfyriadis$^a$ and
F.M. Renard$^b$}\\
\vspace{0.7cm}
$^a$Department of Theoretical Physics, University of Thessaloniki,\\
Gr-54006, Thessaloniki, Greece.\\
\vspace{0.2cm}
$^b$Physique
Math\'{e}matique et Th\'{e}orique,
UMR 5825\\
Universit\'{e} Montpellier II,
 F-34095 Montpellier Cedex 5.\\
\vspace{0.2cm}

\vspace*{1cm}

{\bf Abstract}
\end{center}
We point out a few remarkable properties of the
$\gamma\gamma\to\gamma\gamma$ process at high energy, which
should allow to search for effects of new particles and
interactions. We give illustrations with threshold effects due to
pairs of new charged particles (charginos, charged Higgs particles,
sfermions), resonance effects due to $s$-channel production of
neutral scalars (standard or supersymmetric neutral Higgs
particles or technipions) and unitarity
saturating amplitudes due to a strongly interacting sector. The use
of polarized photon beams is also briefly discussed.   \par

\def\thefootnote{\arabic{footnote}}
\setcounter{footnote}{0}
\clearpage

In this letter we point out a few  striking  properties
of the $\gamma\gamma \to \gamma\gamma$ scattering amplitude
at high energy, which  may be very useful
in  the search of new particles and
interactions in a  $e^-e^+$ Linear Collider
(LC) \cite{LC} operated in the $\gamma \gamma $ mode \cite{LCgg}.
We find that for such an LC, $\gamma\gamma \to \gamma\gamma$ scattering
may be a very useful tool for new physics  searches. \par

The invariant helicity amplitudes
$F_{\lambda_1 \lambda_2 \lambda_3\lambda_4}$ for  the process
\bq
\gamma (p_1,\lambda_1) \gamma (p_2,\lambda_2) \to
\gamma (p_3,\lambda_3) \gamma (p_4,\lambda_4) \ \ ,
\label{gggg-process}
\eq
(where the momenta and helicities of the various photons are
indicated in parenthesis), satisfy
$F_{\lambda_1 \lambda_2 \lambda_3\lambda_4}=
F_{-\lambda_1-\lambda_2- \lambda_3-\lambda_4}$
because of parity conservation, and are determined
in the Standard Model (SM)  by 1-loop diagrams
involving contributions from charged fermions
\cite{vanderBij} and $W^{\pm}$ bosons \cite{Jikia}.
For $\sqrt{s_{\gamma \gamma}} \gsim 250 GeV$, the  basic features
of these amplitudes are the dominance of the
$W$ contribution over the quark and
leptonic ones; and the fact that only the amplitudes
\bqa
& F_{\pm\pm\pm\pm}(\s,\t,\u) & \ \ , \nonumber \\
& F_{\pm\mp\mp\pm}(\s,\u,\t) =F_{\pm\mp\pm\mp}(\s,\t,\u) & \ ,
\label{pmmp}
\eqa
are appreciable, where  $\s=(p_1+p_2)^2$, $\t=(p_3-p_1)^2$,
$\u=(p_4-p_1)^2$, (compare \ref{gggg-process}). \par

The additional
remarkable fact is that these dominant amplitudes
are almost purely imaginary at such energies,
for all scattering angles, small\footnote{In effect, the
amplitude in (\ref{gggg-process}) has the $s$-channel helicity
conservation properties anticipated long ago on the basis of
Vector Meson Dominance and Pomeron exchange. But of course
the role of the Pomeron is now taken by the $W$-loop, which is
not connected with the Pomeron in an obvious way.}
\cite{Jikia} and large \cite{GPRsusy}.
This is illustrated in Fig.1a,b \cite{GPRsusy}.
Using this and the expressions in \cite{vanderBij, Jikia, GPRsusy},
we find that the
dominant amplitudes for $\hat{s},|\hat{t}|,|\hat{u}|
\gg M^2_W$, are
\bqa
 F_{\pm\pm\pm\pm}(\hat{s},\hat{t},\hat{u})
& \simeq &  -i~ 16\pi\alpha^2\Bigg [{\hat{s}\over \hat{u}}
Ln\Big |{\hat{u}\over M^2_W}\Big | +
{\hat{s}\over \hat{t}} Ln\Big |{\hat{t}\over M^2_W}\Big |\Bigg ]
\ \ , \label{pppp} \\
 F_{\pm\mp\pm\mp}(\hat{s},\hat{t},\hat{u})
& \simeq &
-i~ 12\pi\alpha^2{\hat{s}-\hat{t}\over \hat{u}}+i{8\pi\alpha^2\over
\hat{u}^2}(4\hat{u}^2-3\hat{s}\hat{t})
\Bigg [Ln\Big |{\hat{t}\over \hat{s}}\Big |\Bigg ]\nonumber\\
&&  -i~ 16\pi\alpha^2 \Bigg [{\hat{u}\over \hat{s}}
Ln \Big |{\hat{u}\over M^2_W}\Big | + {\hat{u}^2\over \hat{s}\hat{t}}
Ln \Big |{\hat{t}\over M^2_W} \Big |\Bigg ]
\label{pmpm} \ ,
\eqa
which  imply that at $\vartheta^*=90^{\underline{0}}$
 $ F_{\pm\pm\pm\pm}\simeq 4
F_{\pm\mp\pm\mp} \simeq 4  F_{\pm\mp\mp\pm}$ \cite{GPRsusy}. The
same expressions (\ref{pmmp}-\ref{pmpm}) imply that
 at small angles  $ F_{\pm\pm\pm\pm} \simeq F_{\pm\mp\pm\mp}$, while
 $F_{\pm\mp\mp\pm}$ is negligible \cite{Jikia}; as it is also
confirmed by the results of the exact computation
shown in Fig.1b.

We now propose to use  these properties for searching for
new physics phenomena (NP), through precision measurements of the
$\gamma\gamma\to\gamma\gamma$ cross section. An NP amplitude
(too weak to show up through its modulus squared) will
contribute through its interference with the SM amplitude.
Because of the dominance of the imaginary parts in the SM
amplitudes, the only significant
interference terms will  pick up the imaginary part of the new
amplitude. This  is particularly appealing, as it  allows
to observe new physics effects
characterized by amplitudes with a large imaginary part;
like  e.g. threshold effects due to the
virtual production  of pairs of new particles;
resonant contributions  due
to the formation of neutral states coupled to $\gamma\gamma$;
and unitarity saturating amplitudes due to a strongly interacting
scalar sector.\par

Before developing these, let us say a few words about the
experimental aspects of the $\gamma \gamma$ collision process,
which may be studied at  an $e^+e^-$ linear collider (LC) through
the laser backscattering method \cite{LC, LCgg}.
This method should lead to
$\gamma\gamma$ collisions  at energies
$E^{\gamma\gamma}_{cm} \lsim 0.8~ E^{ee}_{cm}$ with a luminosity
close or even larger than the $e^+e^-$  one
$\L_{ee}$,  depending on the tuning of the backscattering
process. The presently contemplated value  for the LC project \cite{LC} is
$\L_{ee} \simeq 500~-~1000 fb^{-1}$ per one or two
years of running in \eg\@ the high luminosity TESLA mode at
 energies of $350-800 GeV$.
Since the $\gamma \gamma$ cross section is
$\bar \sigma (\gamma\gamma\to\gamma\gamma)\simeq 10fb$,
for a broad range of $\hat{s}\equiv s_{\gamma\gamma}$
at large c.m. angles
($30^{\underline{0}}<\vartheta^*<150^{\underline{0}}$)
\cite{Jikia, GPRsusy}, it should possible to
collect several thousands of $\gamma\gamma\to\gamma\gamma$
events in this large energy and  angle range.
In this domain the measurements  should
be very clean and the background negligible.\par

Returning now to the use of $\gamma \gamma \to \gamma \gamma$
for NP searching, we first discuss the unpolarized case, for which the
cross section  reads
\bqa
{d\bar \sigma_0\over d\cos\vartheta^*}&=&
\left ({1\over128\pi\hat{s}}\right )
\sum_{\lambda_3\lambda_4} [|F_{++\lambda_3\lambda_4}|^2
+|F_{+-\lambda_3\lambda_4}|^2]\nonumber\\
&&\simeq \left ({1\over128\pi\hat{s}}\right )
[(ImF_{++++})^2+(Im F_{+-+-})^2+(Im F_{+--+})^2] \ .
\label{sig0}
\eqa
Writing $F=F^{SM}+F^{new}$, the dominant contribution containing
the new amplitude will be
\bqa
{d\bar \sigma_0\over d\cos\vartheta^*} & \simeq &
\left ({1\over64\pi\hat{s}}\right )[Im F^{SM}_{++++}Im
F^{new}_{++++} \nonumber \\
&& +Im F^{SM}_{+-+-}Im F^{new}_{+-+-}+Im
F^{SM}_{+--+}Im F^{new}_{+--+}] \ \ .
\label{signew}
\eqa

As a first illustration we consider the contribution
to $\gamma\gamma\to\gamma\gamma$ from
a box loop in which a new charged particle $X^\pm$ circulates.
$X^{\pm}$ may be  a fermion, (like \eg\@  a chargino), or
a charged scalar (like a charged Higgs, a sfermion or a
technipion). The
imaginary part of the new amplitudes can  obtained from
the expressions given in \cite{vanderBij, Jikia} and reproduced in
\cite{GPRsusy},
in terms of the
of the   1-loop $B,C,D$ functions, using the imaginary parts
\bq
Im B(\hat{s})=\pi\sqrt{1-{4M^2_X\over\hat{s}}}~~\theta(\s-4M^2_X)
\ ,
\eq
\bqa
Im C(\hat{s})& =& -{2\pi\over \hat{s}}
Ln \Bigg |\sqrt{{\hat{s}\over4M^2_X}} \left
(1+\sqrt{1-{4M^2_X\over\hat{s}}}\right )\Bigg | \theta(\s-4M^2_X)
\ \ , \\
Im D(\hat{s},\hat{t})& = & {\pi\over
\sqrt{\hat{s}^2\hat{t}^2+4M^2_X\hat{s}\hat{t}\hat{u}}}
Ln\Bigg |{(x^+-x_a)(x^--x_b)\over(x^--x_a)(x^+-x_b)}
\Bigg |\theta(\s-4M^2_X) \ ,
\label{imD}
\eqa
with
\bq
x^{\pm}={1\over2}\left [1\pm\sqrt{1-{4M^2_X\over\hat{s}}}\right ]
\ \ \ , \ \ \
x_{a,b}={1\over2}\left [-{\hat{t}\over \hat{u}}\pm\sqrt{{\hat{t}^2
\over \hat{u}^2}
+{4M^2_X\hat{t}\over \hat{s}\hat{u}}}\right ]\ \ .
\eq

Thus for a fermion $X^\pm$-particle loop,
we obtain
\bqa
&&Im F^{f}_{\pm\pm\pm\pm}(\hat{s},\hat{t},\hat{u})=
8\alpha^2 Q_X^4  M^2_X(s-2M^2_X)[Im D(\hat{s},\hat{t})+Im
D(\hat{s},\hat{u})] \ , \nonumber \\
&&Im F^{f}_{\pm\pm\mp\mp}(\hat{s},\hat{t},\hat{u})
=-16\alpha^2 Q_X^4  M^4_X[Im D(\hat{s},\hat{t})+Im
D(\hat{s},\hat{u})] \ ,\nonumber\\
&&Im F^{f}_{\pm\mp\pm\mp}(\hat{s},\hat{t},\hat{u})=
Im F^{f}_{\pm\mp\mp\pm}(\hat{s},\hat{u},\hat{t})=
\alpha^2 Q_X^4 \cdot \Bigg \{
 8\left ({\hat{s}-\hat{t}\over \hat{u}}\right )Im B(s) \nonumber \\
&& -8\left [{t^2+s^2-4M^2_X \hat{u}\over \hat{u}^2}\right ]
\hat{s} Im C(\hat{s})
+8M^2_X(\hat{u}-2M^2_X)Im D(\hat{s},\hat{u}) \nonumber\\
&& +{4\over
\hat{u}^2}[-4M^2_X \hat{u}(M^2_X\hat{u}+\hat{t}\hat{s})+
(2\hat{u}M^2_X+\hat{t}\hat{s})(\hat{t}^2+\hat{s}^2)]
Im D(\hat{s},\hat{t}) \Bigg \} \ .
\label{newferm}
\eqa
Close to the threshold $\hat{s} \sim 4M^2_X$, and at
$\vartheta^* \sim 90^{\underline{0}}$,
the dominant contributions from (\ref{newferm}) are then
\bq
Im F^{f}_{\pm\pm\pm\pm}\simeq-Im F^{f}_{\pm\pm\mp\mp}\simeq
 8\pi\alpha^2 Q_X^4 \sqrt{1-{4M^2_X\over\hat{s}}}~~~
\theta(\hat{s}-4M^2)
\label{th}
\eq
\noindent
whereas the other terms $Im F^{f}_{\pm\mp\pm\mp}$,
$Im F^{f}_{\pm\mp\mp\pm}$ are of higher order with respect
to the small quantity  $\sqrt{1-4M^2_X/\s}$.\par

The interference of these amplitudes with the imaginary parts of
the SM ones induces a threshold effect, giving a clear
signal for the production of the $X^\pm$ particles. This is also
confirmed by the exact computation  shown in Fig.2a,
in which the real and imagianry parts of all amplitudes are
retained \cite{GPRsusy}.   Thus,  the unpolarized cross
section reflects in a perfect way the threshold effect due to
the behavior of the imaginary part of the new amplitudes, without any
appreciable perturbation due to the real parts.
Quantitatively the threshold effect decreases from about
10\% to 5\%,  when $M_X$ increases from $100~GeV$ to
$250~GeV$. This should  be observable at
$\gamma\gamma$ collisions obtained through laser backscattering
at an LC collider of $E_{ee}=800~GeV$.
Qualitatively similar, but quantitatively considerably smaller
predictions  are also obtained for the box contribution from a
scalar charged particle, \cite{GPRsusy}.\par

On the basis of the results of Fig.2a, it is also worthy to remark
that the use of $\gamma \gamma \to \gamma  \gamma$ scattering
in order  to search for  the presence of new $X^\pm$ particles,
avoids  the difficult
task of studying their decay modes, in case  they are  actually
produced.  Such decay mode studies are indeed
often hindered by  huge backgrounds and they are also  affected by
(possibly many) new
parameters. This is \eg\@  the  case  in the SUSY
models, in which  the  new sparticle decays
are affected by  a large number of  symmetry breaking parameters
\cite{GPRsusy}. \par

As a second illustration we consider a single neutral scalar
$X^0$ of mass $M_X$, which  could be a neutral Higgs boson,
a Technipion or any other neutral scalar state.
The imaginary parts of the new (resonant) amplitudes are
\bq
Im F^{X^0}_{\pm\pm\pm\pm}=Im F^{X^0}_{\pm\pm\mp\mp}=16\pi{M_X^2
\Gamma_{X\to\gamma\gamma} \Gamma_X\over
(\hat{s}-M^2_X)^2+M^2_X\Gamma^2_X} \ \ ,
\label{res}
\eq
which substituted in (\ref{signew}) create the effect shown in
Fig.2b when $M_X=400~GeV$, $\Gamma_X=10~GeV$ and
$\Gamma_{X\to\gamma\gamma}=10^{-4}~GeV$. \par

Using  (\ref{signew}), (\ref{pppp})-(\ref{pmmp}) and
(\ref{res}) we see that the observability of a $\gsim 3\%$
effect in $\bar {\sigma_0}$ around $\hat{s} \simeq M^2_X$, requires
a branching ratio $B(X\to\gamma\gamma)\equiv
\Gamma_{X\to\gamma\gamma}/\Gamma_X
\gsim (0.5~-~1)\times10^{-5}$ when $M_X$ lies in the
several hundreds of GeV range. This condition is satisfied in
many cases; like e.g.   for a standard Higgs
with $m_H\lsim 400~GeV$; and for the MSSM Higgs particles $H,~ A$
with masses up to the TeV range  \cite{book,Spira}.
In the technicolour case, things are more model dependent,
\cite{TC}; but usually the $\gamma\gamma$ branching ratio for  a
technipion is much higher than in the Higgs case, so that one can
expect an even clearer effect; see \eg\@ \cite{TC1} where
the enormous values of $10^{-2}$ to $10^{-3}$ are quoted.\par

As a last example we  quote the case of a new amplitude due
to the formation of a broad scalar or due to the unitarity
saturating mechanism associated to a strongly interacting
sector \cite{strong}. Such an amplitude has the tendency of being purely
imaginary, as one can see from a broad Breit-Wigner expression
or more generally from a diffractive picture generalizing the
$VV\to VV$ hadronic case. It would then also directly show up in
$\gamma\gamma\to \gamma\gamma$ according to eq.(\ref{signew}).
The effect
could start at the threshold by the  production of the lightest pair
of charged particles (like e.g charged technipions), and it may stand
up to the highest available energy, through the addition of the
full spectrum of states associated to this new sector. Writing
as a simple ansatz
\bq
Im F^{sat}_{\pm\pm\pm\pm}=Im F^{sat}_{\pm\pm\mp\mp}=16\pi
B_{\gamma\gamma}
\sin^2\left ({\pi\over2}(1-{4M^2_P\over\hat{s}})^{1/2}\right )
\label{sat}\eq
with $M_P=200~GeV$ and $B_{\gamma\gamma}=10^{-4}$, one obtains
the effect depicted in Fig.2c.\par

Let us add a short discussion on the possible use of polarized
$\gamma\gamma$ collisions in an LC operated in the $\gamma
\gamma $ mode. With Parity invariance and Bose statistics
the general form of the polarized $\gamma\gamma$
cross section in an LC  simplifies to \cite{Tsi}, \cite{GPRsusy}
\bqa
{d\sigma\over d\tau d\cos\theta^*}&=&{d\L_{\gamma\gamma}\over
d\tau} \Bigg \{
{d\bar{\sigma}_0\over d\cos\theta^*}
+<\xi_2\tilde \xi_2>{d\bar{\sigma}_{22}\over d\cos\theta^*}
+[<\xi_3>\cos2\phi+<\tilde\xi_3>\cos2\tilde\phi]
{d\bar{\sigma}_{3}\over d\cos\theta^*}
\nonumber\\
&&+<\xi_3\tilde \xi_3>[{d\bar{\sigma}_{33}\over d\cos\theta^*}
\cos2(\phi+\tilde\phi)
+{d\bar{\sigma}^\prime_{33}\over
d\cos\theta^*}\cos2(\phi-\tilde\phi)]\nonumber\\
&&+[<\xi_2\tilde\xi_3>\sin2\tilde\phi-<\xi_3\tilde \xi_2>\sin2\phi]
{d\bar{\sigma}_{23}\over d\cos\theta^*} \Bigg \} \ \ ,
\label{sigpol}\eqa
where $d\bar{\sigma}_0/d\cos\theta^*$ is given in (\ref{sig0}),
and
\bqa
{d\bar{\sigma}_{22}\over d\cos\theta^*} &=&
\left ({1\over128\pi\hat{s}}\right )\sum_{\lambda_3\lambda_4}
[|F_{++\lambda_3\lambda_4}|^2
-|F_{+-\lambda_3\lambda_4}|^2]  \ , \label{sig22} \\
{d\bar{\sigma}_{3}\over d\cos\theta^*} &=&
\left ({-1\over64\pi\hat{s}}\right ) \sum_{\lambda_3\lambda_4}
Re[F_{++\lambda_3\lambda_4}F^*_{-+\lambda_3\lambda_4}]  \ ,
\label{sig3} \\
{d\bar \sigma_{33} \over d\cos\theta^*}& = &
\left ({1\over128\pi\hat{s}}\right ) \sum_{\lambda_3\lambda_4}
Re[F_{+-\lambda_3\lambda_4}F^*_{-+\lambda_3\lambda_4}] \ ,
\label{sig33} \\
{d\bar{\sigma}^\prime_{33}\over d\cos\theta^*} &=&
\left ({1\over128\pi\hat{s}}\right ) \sum_{\lambda_3\lambda_4}
Re[F_{++\lambda_3\lambda_4}F^*_{--\lambda_3\lambda_4}] \  ,
\label{sig33prime} \\
{d\bar{\sigma}_{23}\over d\cos\theta^*}& = &
\left ({1\over64\pi\hat{s}}\right ) \sum_{\lambda_3\lambda_4}
Im[F_{++\lambda_3\lambda_4}F^*_{+-\lambda_3\lambda_4}] \ .
\label{sig23}
\eqa

In (\ref{sigpol}) the quantity $d\L_{\gamma\gamma}/d\tau$
 describes the overall photon-photon luminosity
(per unit $e^-e^+$ flux) in an LC operated in the $\gamma \gamma$ mode
\cite{LCgg}, while $\tau \equiv s_{\gamma \gamma}/s_{ee}$.
The Stokes parameters $\xi_2$ and $-\xi_3e^{-2i\phi}$ in
(\ref{sigpol}), determine the normalized helicity density matrix
of one of the backscattered photons
 $\rho^{BN}_{\lambda \lambda^\prime}$  through \cite{Tsi}
\bq
\xi_2=\rho^{BN}_{++} -\rho^{BN}_{--} \ \ \ , \ \ \
-\xi_3e^{-2i\phi}=\rho^{BN}_{+-} \ .
\eq
Thus, $\xi_2$ describes the average longitudinal
polarization of the backscattered photon, while $\xi_3$
determines the magnitude of the
transverse linear polarization, whose direction is defined
by the azimuthal angle $\phi$,
with respect to the photon momentum. Finally $\tilde\xi$ and
$\tilde \phi$ refer to the second
photon. \par

By making measurements with various  choices of
polarizations of both of the  backscattered photons (which is 
achieved by
suitable choices of the $e^\pm$ and laser photon polarizations),
all five quantities in (\ref{sig22}-\ref{sig23}) can be determined.
The standard (SM) contributions to these (integrated in the range
 $30^{\underline{0}}<\vartheta^*<150^{\underline{0}}$) are plotted in
Fig.3. At large energy the results are exactly the ones that
one expects from the amplitudes given
in eq.(\ref{pppp}-\ref{pmmp}) and Fig.1; \ie\@
\bq
{\bar{\sigma}_{22}\over\bar{\sigma}_0}\simeq0.5 ~~,~~
{\bar{\sigma}_{3}\over\bar{\sigma}_0}\simeq{0.01}~~,~~
{\bar{\sigma}_{33}\over\bar{\sigma}_0}\simeq{0.1}~~,~~
{\bar\sigma^\prime_{33}\over\bar{\sigma}_0}\simeq{0.02}~,~~~
{\bar{\sigma}_{23}\over\bar{\sigma}_0}\simeq{0.01}~~ .
\label{pol}
\eq

When new physics  amplitudes like those in
eq.(\ref{newferm},\ref{res},\ref{sat}) are added,
then the  dominance in the SM part of the purely
imaginary amplitudes listed in eq.(\ref{pppp}-\ref{pmmp}),
leads us to  expect
about the same type of \underline{relative}
effects in $\bar{\sigma}_{22}$,
$\bar{\sigma}_{33}$, $\bar{\sigma}_{23}$
and  in $\bar{\sigma}_0$. However, much larger
\underline{relative} effects are expected for the $\bar{\sigma}_{3}$
and  $\bar \sigma^\prime_{33}$ cases; since  there the SM
contribution is depressed, while the new physics contribution is
not.  For the thresholds and resonance examples illustrated
above and the chosen  values of the parameters, one  may thus expect
a $50-100\%$ effect in the $\bar{\sigma}_{3}$ or
$\bar \sigma^\prime_{33}$ cases, whereas it was $5-10\%$
in the $\bar{\sigma}_0$ one. This
enhancement compensates the fact that the number of events in
$\bar{\sigma}_{3}$ and $\bar{\sigma}_{33}$ is reduced by a factor
$50-100$. In addition a sign difference appears
between the case of a new fermion and the case  of a new scalar.
This should be
observable and extremely useful for the identification of the origin
of an effect;  \eg\@ for disentangling a fermionic threshold
from a bosonic one, leading to unitarity saturation (Fig.2a and 2c).
More detailed analyses will be given elsewhere, \cite{GPRsusy}.\par

In conclusion, with these few illustrations we have shown that
indeed  the $\gamma\gamma\to\gamma\gamma$ process provides
a clean way to check for the presence of new particles and interactions.
This is due to the unique property of this process to have the
SM contribution appearing only at the 1-loop level and to be dominated
by few, purely imaginary, helicity amplitudes. As a
consequence, the presence of new particles or interactions
occurring at the same level (boxes of  charged particles or resonant
single neutrals), would immediately lead to a clear signal.
This method is independent and complementary to the one of
looking at the direct production of new particles and studying
their  decay modes.
It  should be especially advantageous for the search of
new particles decaying through a long chain of processes, which are
difficult to extract from a background. In addition the
availability of polarized $\gamma\gamma$ collisions should
give informations about the nature of the particles produced.\par

\underline{Acknowledgments}\\
We like to thank D. Papadamou and G. Tsirigoti for help in
the early stages of this work.

\clearpage
\newpage

\begin{figure}[p]
\vspace*{-4cm}
\[
\epsfig{file=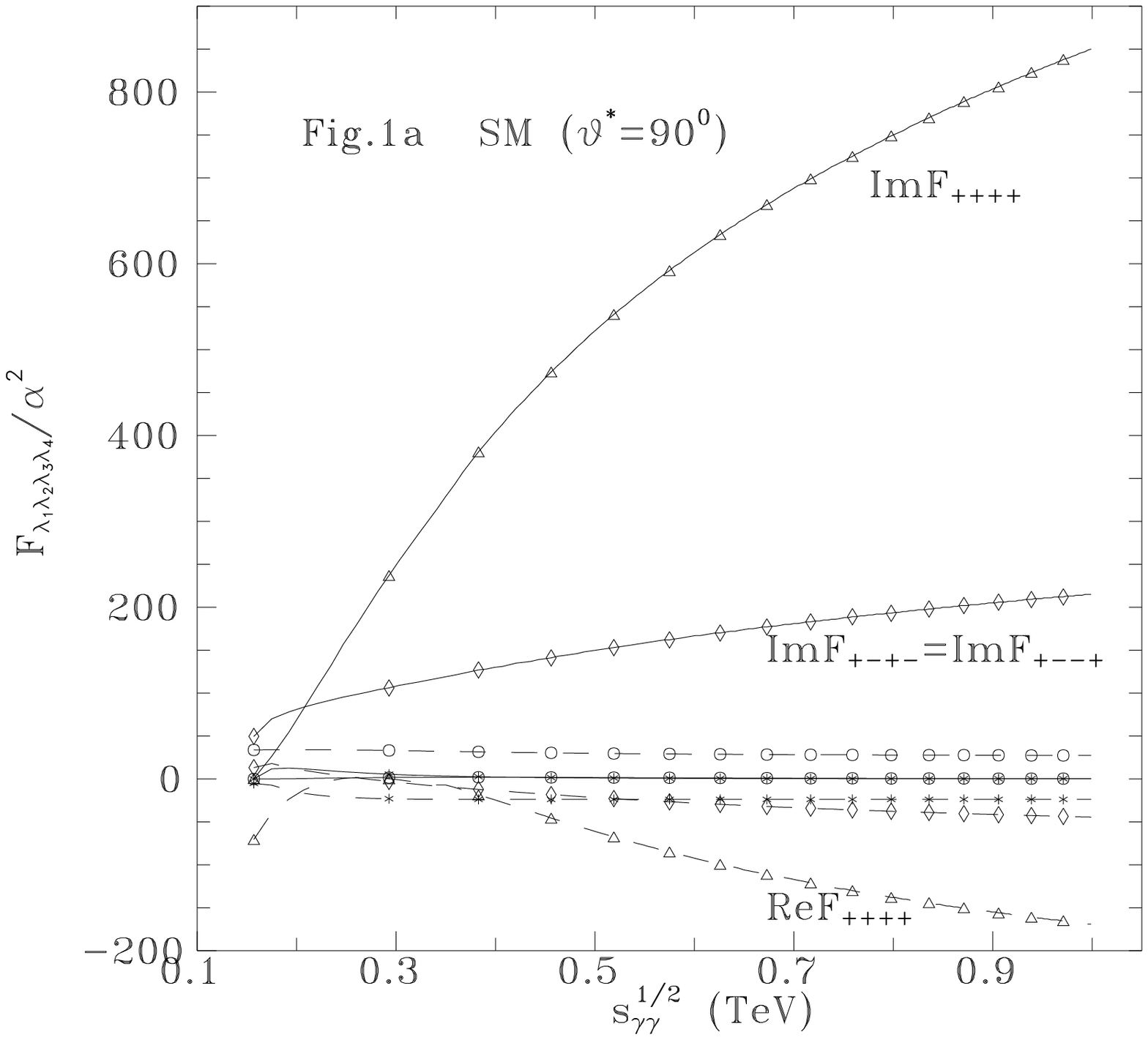,height=7.5cm}\hspace{0.5cm}
\epsfig{file=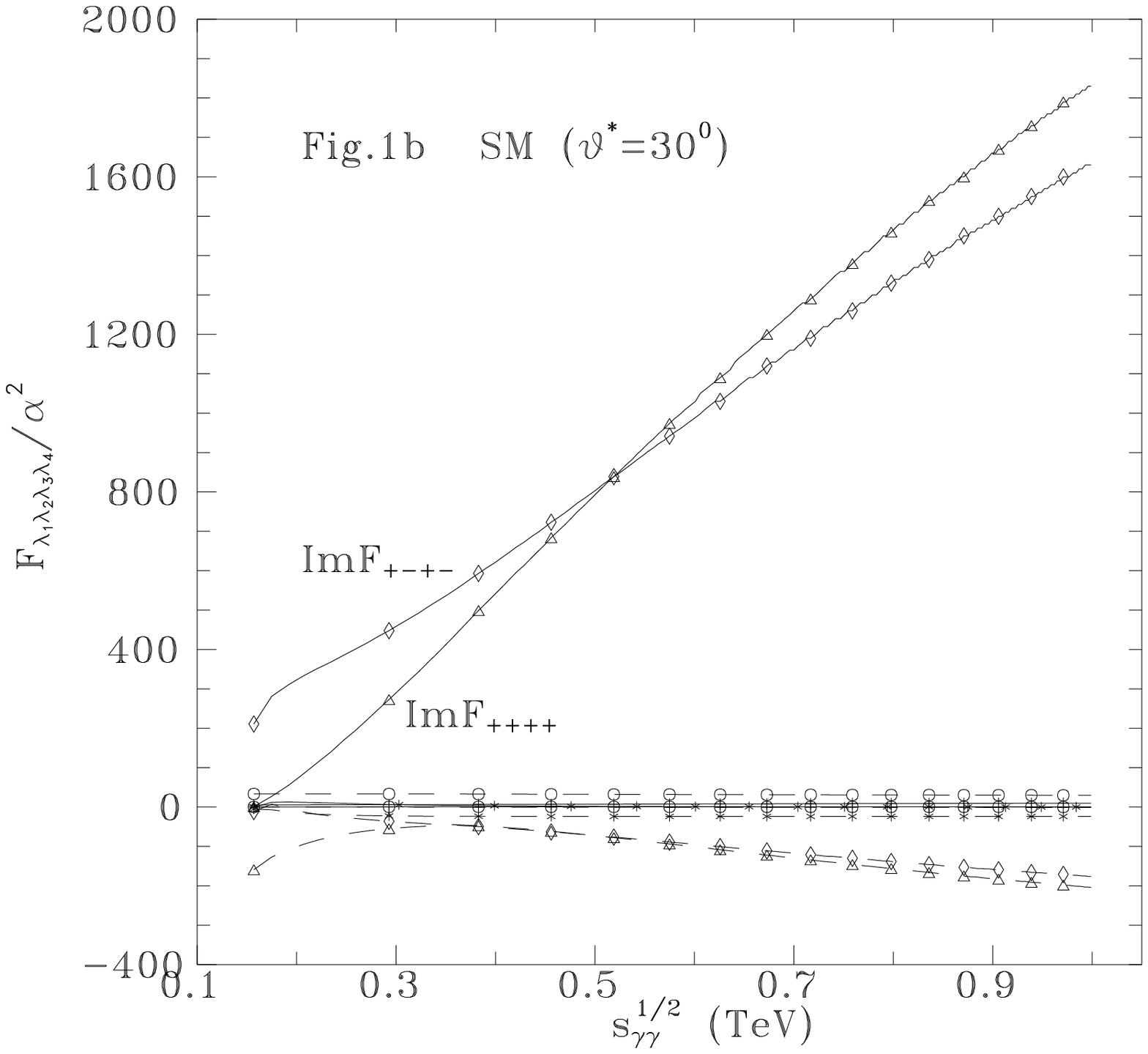,height=7.5cm}
\]
\vspace*{1.cm}
\caption[1]{Imaginary (solid line) and real (dashed line) parts
of the SM $\gamma \gamma \to \gamma \gamma $ helicity amplitudes
at $\vartheta =90^0$ (1a) and $\vartheta =30^0$ (1b)
for $F_{++++}$ (triangles),
$F_{+++-}$ (circles), $F_{++--}$ (stars),
$F_{+-+-}$ (rhombs). Finally $F_{+--+}$ (crosses)
is omitted in the (1a) case, since it is identical to $F_{+-+-}$
at $\vartheta =90^0$, while in the (1b) case it is negligibly small.
We also note that $F_{++-+}(\s,\t,\u)=F_{+-++}(\s,\t,\u)=
F_{+---}(\s,\t,\u)=F_{+++-}(\s,\t,\u)$.}

\label{Figure1ab}
\end{figure}

\clearpage
\newpage

\begin{figure}[p]
\vspace*{-4cm}
\[
\epsfig{file=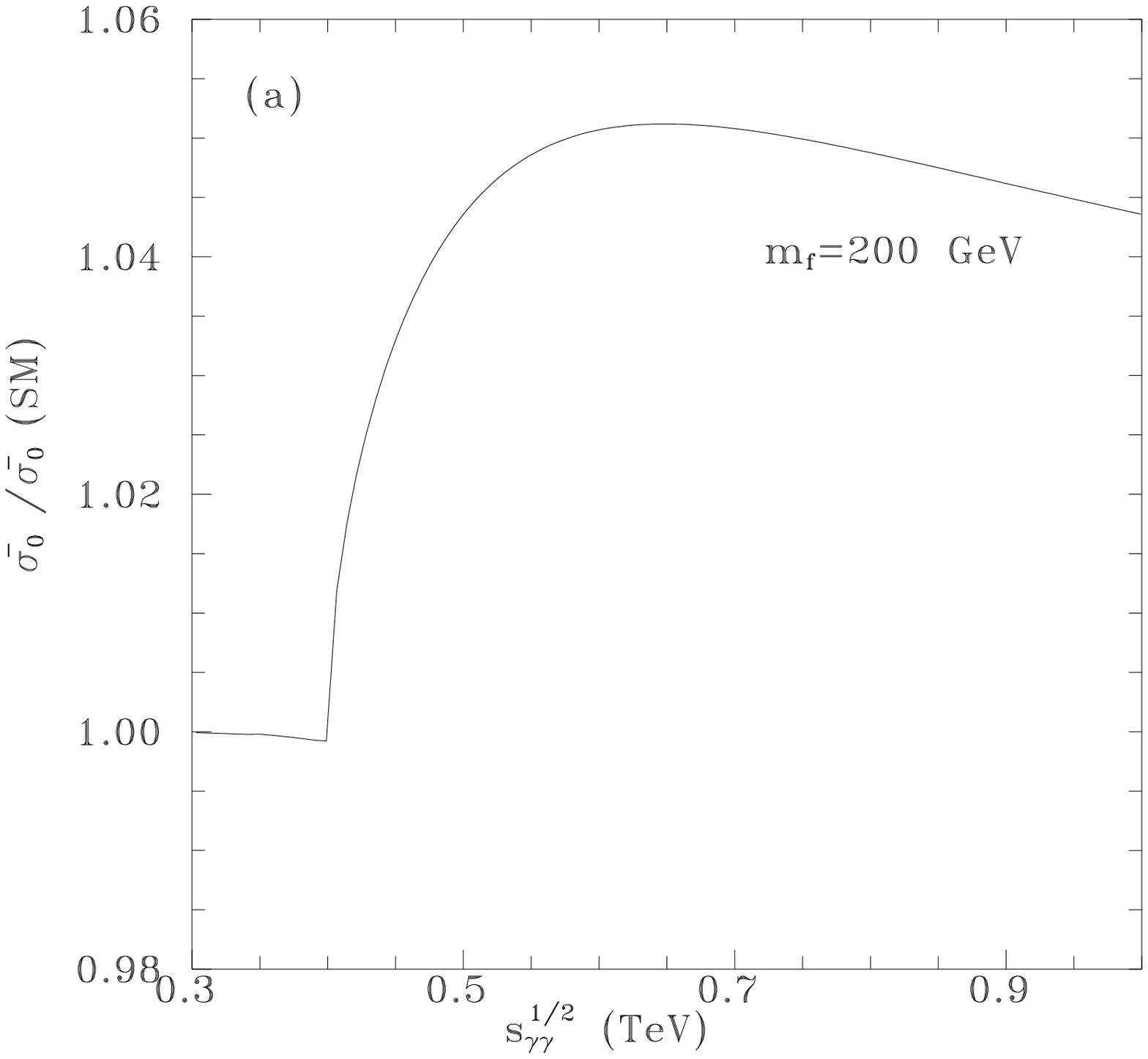,height=7.5cm}\hspace{0.5cm}
\epsfig{file=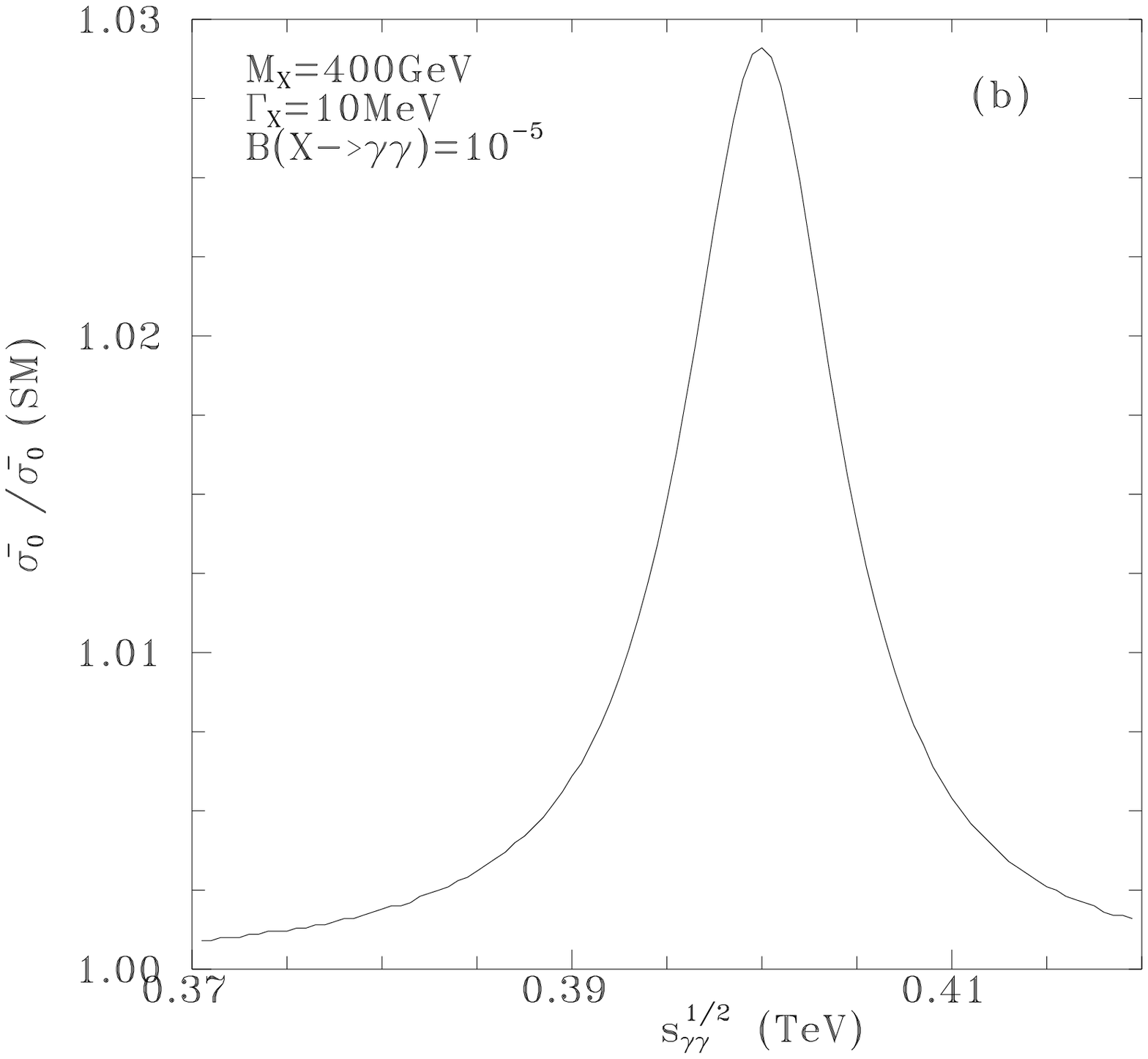,height=7.5cm}
\]
\vspace*{1.5cm}
\[
\epsfig{file=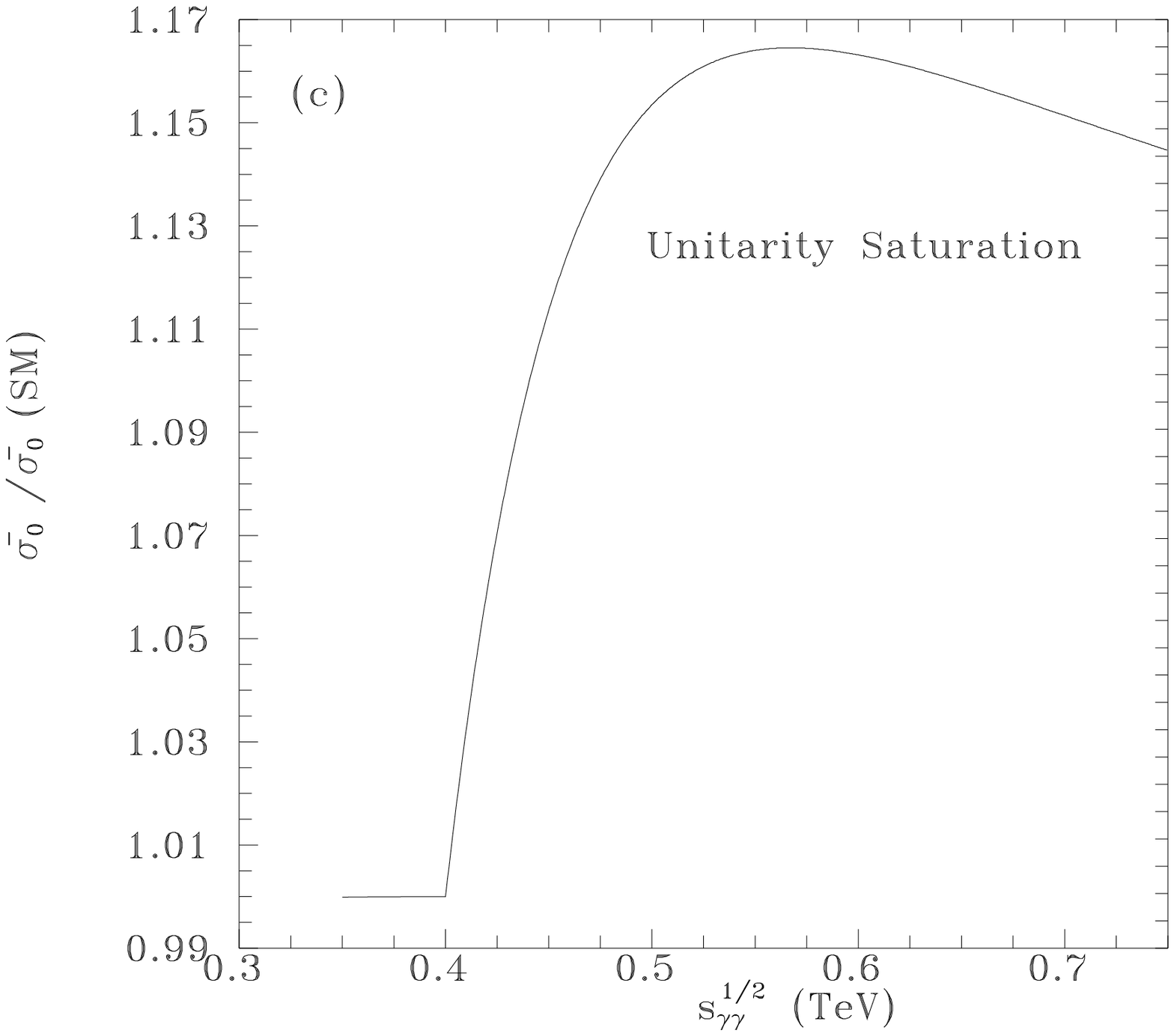,height=7.5cm}
\]
\vspace*{1.cm}
\caption[1]{Relative magnitude with respect to the SM results,
of the unpolarized $\gamma \gamma \to \gamma \gamma $
cross sections for a charge $+1$ fermion (a),
a typical s-channel neutral resonance (b), and 
unitarity saturating amplitudes (c). In all cases 
the cross sections have been integrated in the c.m. angular range
$30^0 < \vartheta^* < 150^0$. }
\label{models}
\end{figure}

\clearpage
\newpage

\begin{figure}[p]
\vspace*{-4cm}
\[
\epsfig{file=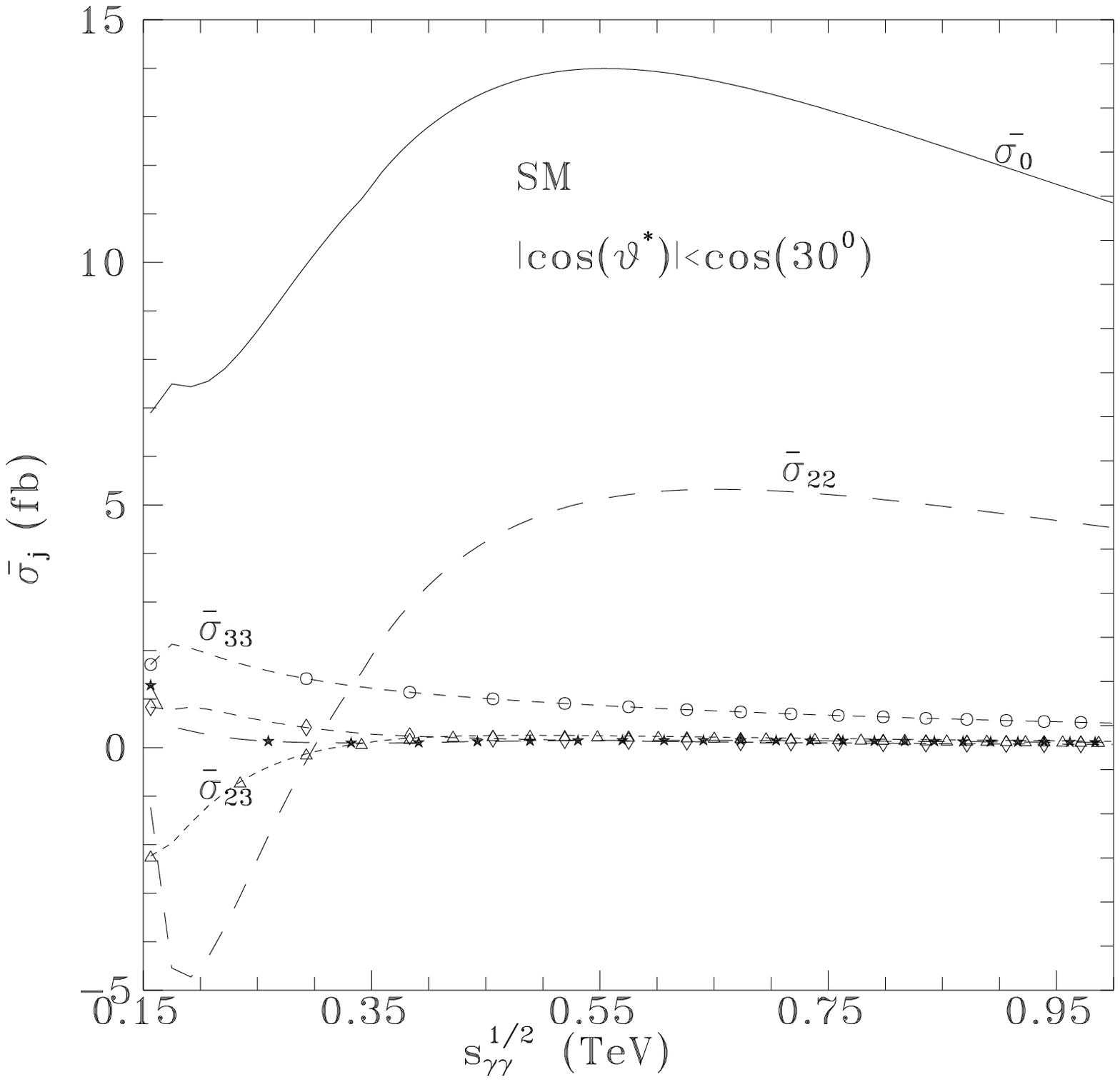,height=15cm}
\]
\vspace*{1.cm}
\caption[1]{ SM predictions for $\bar \sigma_0$ (solid),
$\bar \sigma_{22}$ (dash), $\bar \sigma_3$ (stars)
$\bar \sigma_{33}$ (circles), $\bar \sigma_{33}^\prime$ (rhombs),
$\bar \sigma_{23}$ (triangles). } 
\label{Fig-sigsm}
\end{figure}


\newpage





\begin{thebibliography}{99}
%
\bibitem{LC} Opportunities
and Requirements for Experimentation at a Very High Energy
$e^{+}e^{-}$ Collider, SLAC-329(1928); Proc. Workshops on Japan
Linear Collider, KEK Reports, 90-2, 91-10 and 92-16;
P.M. Zerwas, DESY 93-112, Aug. 1993; Proc. of the Workshop on
$e^{+}e^{-}$ Collisions at 500 GeV: The Physics Potential, DESY
92-123A,B,(1992), C(1993), D(1994), E(1997) ed. P. Zerwas;
E. Accomando \etal\@ \prep{C299}{1998}{299}.
%
\bibitem{LCgg} I.F. Ginzburg, G.L. Kotkin, V.G. Serbo
and V.I. Telnov, Nucl. Instr. and Meth. {\bf 205}, (1983) 47;
I.F. Ginzburg, G.L. Kotkin, V.G. Serbo, S.L. Panfil and V.I. Telnov,
Nucl. Instr. and Meth. {\bf 219},(1984) 5; J.H. K\"{u}hn, E.Mirkes
and J. Steegborn, \zp{C57}{1993}{615}.
%
\bibitem{vanderBij} E.W.N. Glover and J.J. van der Bij
\np{B321}{1989}{561}.
%
\bibitem{Jikia} G. Jikia and A. Tkabladze, \pl{B323}{1994}{453}.
%
\bibitem{GPRsusy} G.J. Gounaris, P.I. Porfyriadis F.M.
Renard and G. Tsirigoti, in preparation.
%
\bibitem{book} J.F. Gunion, H.E. Haber, G. Kane and S.
Dawson, The Higgs Hunter's Guide, Adison-Wesley, Redwood City,
Ca, 1990.
%
\bibitem{Spira} M. Spira, hep-ph/9705337, Fortsch. Phys.
{\bf 46} (1998) 203.
%
\bibitem{TC} K. Lane and E. Eichten, \pl{B352}{1995}{382};
\pr{D54}{1996}{2204}.
%
\bibitem{TC1} R. Casalbuoni et al., hep-ph/9809523.
%
\bibitem{strong} M. Chanowitz and M.K. Gaillard, \np{B261}{1985}{379}.
%
\bibitem{Tsi} G.J. Gounaris and G. Tsirigoti,
\pr{D56}{1997}{3030}, Erratum-ibid {\bf D58} (1998) 059901.
%
\end{thebibliography}
\end{document}